\documentclass[11pt,a4paper]{article}
\usepackage{jheppub}
\usepackage{pifont}
\usepackage{appendix}

\newcommand\fverb{\setbox\fverbbox=\hbox\bgroup\verb}
\newcommand\fverbdo{\egroup\medskip\noindent%
            \fbox{\unhbox\fverbbox}\ }
\newcommand\fverbit{\egroup\item[\fbox{\unhbox\fverbbox}]}
\newbox\fverbbox

\title{Parametrization of degenerate density matrices}
\author{Tae-Hun Lee}
\emailAdd{taehunee@gmail.com}

\affiliation{School of Mathematical Sciences, University of KwaZulu-Natal, Westville, Durban 4000, South Africa}

\abstract{This paper presents a parametrization of a degenerate density matrix. The problem needs to be approached first with a diagonalized form (the spectral representation) to deal with degeneracy. Such a form is useful for this parametrization in that the conditions to be a density matrix from a Hermitian matrix are applied only to a diagonal eigenvalue matrix, not a unitary matrix. Those conditions can be satisfied by parametrizing eigenvalues with squared spherical coordinates in dimension of the matrix.\\ 
Degeneracy in eigenvalues brings symmetries between a eigenvalue matrix and a unitary matrix, which are realized in a form of a commuting unitary matrix, called a commutant. The associated redundant parameters in a unitary matrix have to be eliminated.\\
It is realized in this paper that degrees of degeneracies can be defined as the total number of possible pairs of the same eigenvalues and one degree of degeneracy corresponds to one phase and one two dimensional rotation in a unitary matrix or a commutant. In this way all the degeneracies are identified and assigned to one phase-one rotation block. Therefore, a unitary matrix or a commutant is a product of these blocks and a general diagonal phase matrix.\\
In physics a unitary matrix is often parametrized by rotation and phase matrices, called an angular representation here. There are many possible different phase configurations. It is often not a trivial matter whether a given parametrized unitary matrix is general. A simple diagram will be introduced to illustrate how to transform one phase configuration to another.}


\dedicated{}

\begin{document}
\maketitle
\section{Introduction}
A density matrix is a Hermitian matrix satisfying the conditions, the normalized trace and non-negativity of eigenvalues
\begin{equation}
\text{Tr}(\rho)=1,~~~ \lambda_i\geq 0,
\end{equation}
where $\rho$ is a density matrix and $\lambda_i$ are its eigenvalues.
These conditions along with the hermiticity specify relations among elements of the matrix, i.e. a density matrix needs to be parametrized. Moreover, some specific physical situations may impose additional conditions. This presentation is specially interested in cases that some eigenvalues of the matrix are degenerate. In this case first a diagonalized form of the matrix need be considered (the spectral representation). 
\begin{equation}
\rho=UDU^\dagger.
\end{equation}
When a diagonal eigenvalue matrix $D$ is degenerate, the number of independent parameters not only for an eigenvalue matrix but also a unitary matrix $U$ must be reduced from those for a non-degenerate case. A commutant is a mathematical object to express the symmetries occurring due to degeneracy, which is defined as a commuting unitary matrix with an eigenvalue matrix. In general, a diagonal phase matrix is a commutant for any Hermitian matrix, regardless of degeneracy. However, when some of eigenvalues are equal, more matrices in addition to a diagonal phase matrix can commute with a diagonal eigenvalue matrix. Therefore, if a unitary matrix can be parametrized for a commutant to be factored out from it, such a part in a unitary matrix can be eliminated. 
The purpose of this paper is to find a systematic way to get rid of redundant parameters in a general $n$-dimensional degenerate density matrix. For a  density matrix when two eigenvalues are equal, $\lambda_i=\lambda_j$, a possible commutant other than a diagonal phase matrix is a rotation matrix having a $(i,j)$ rotation block. This example suggests that degeneracies could be identified through a separable and factorizable matrix unit with a one-to-one correspondence. If so, it would be some combination of a rotation matrix and phases. Finally, a unitary matrix or a commutant can be constructed as a product of such units and a general phase matrix.\\
The main questions of this paper are following. The first is whether degrees of degeneracies can be practically countable. Second, if so, how are they related with a unitary matrix?  Lastly, what could be a separable matrix unit corresponding to one degree of degeneracy? For practical purposes related to these issue a simple diagram will be introduced how to transform one phase representation to another.

\section{Independent degrees of freedom in a unitary matrix}\label{Sec:Independent degrees of freedom in a unitary matrix}
This section will investigate independent degrees of freedom in a unitary matrix for a degenerate density matrix without using a specific representation. Without loss of generality an example for $n=4$ will be considered, first for instance, for a case  $\lambda_1=\lambda_2$ and $\lambda_3\neq\lambda_4$ and then the other case $\lambda_1=\lambda_2$ and $\lambda_3=\lambda_4$ (Note that the convention $\lambda_1\geq\lambda_2\geq\lambda_3\geq\lambda_4$ is used here.). In the end  it will be realized from this example that it can be a convenient choice, a unitary matrix as a product of rotation matrices with each of them having one phase and a general phase matrix on the right or the left. The following splitting of an eigenvalue matrix is convenient for the purpose. 
\begin{equation}\label{eq:D}
D=\lambda_1\left(\begin{array}{cccc}
1&0&0&0\\
0&1&0&0\\
0&0&1&0\\
0&0&0&1
\end{array}\right)
+
\left(\begin{array}{cccc}
0&0&0&0\\
0&0&0&0\\
0&0&\Delta\lambda_{31}&0\\
0&0&0&\Delta\lambda_{41}
\end{array}\right)
=\lambda_1I+D^\prime,
\end{equation}
where $\Delta\lambda_{31}\equiv\lambda_3-\lambda_1$, $\Delta\lambda_{41}\equiv\lambda_4-\lambda_1$.
The first term $\lambda_1I$ is separated from a unitary matrix and the density matrix can be written,
\begin{equation}\label{eq:density matrix splitting}
\rho=\lambda_1I+
UD^\prime U^\dagger,
\end{equation}
where
\begin{equation}\label{eq:UD1}
U D^\prime=\left(\begin{array}{cccc}
\otimes&\otimes&a_1&b_1\\
\otimes&\otimes&a_2&b_2\\
\otimes&\otimes&a_3&b_3\\
\otimes&\otimes&a_4&b_4
\end{array}\right)
\left(\begin{array}{cccc}
0&0&0&0\\
0&0&0&0\\
0&0&\Delta\lambda_{31}&0\\
0&0&0&\Delta\lambda_{41}
\end{array}\right).
\end{equation}
Irrelevant elements due to the zero entries of the first and second rows in $D^\prime$ are denoted by $\otimes$, which do not contribute to a density matrix. Eq.(\ref{eq:UD1}) is written,
\begin{equation}\label{eq:UD2}
UD^\prime=P_R\left(\begin{array}{cccc}
\otimes&\otimes&|a_1|&e^{i\delta_1}|b_1|\\
\otimes&\otimes&|a_2|&e^{i\delta_2}|b_2|\\
\otimes&\otimes&|a_3|&e^{i\delta_3}|b_3|\\
\otimes&\otimes&|a_4|&|b_4|
\end{array}\right)P_L
\left(\begin{array}{cccc}
0&0&0&0\\
0&0&0&0\\
0&0&\Delta\lambda_{31}&0\\
0&0&0&\Delta\lambda_{41}
\end{array}\right),
\end{equation}
where $P_R$ and $P_L$ are external diagonal phase matrices. $A=(a_1,a_2,a_3,a_4)$ and $B=(b_1,b_2,b_3,b_4)$, where $a_i$ and $b_i$ are complex, are vectors satisfying unitary conditions. The focus here is to count a change of the number of independent internal degrees of freedom due to degeneracy. It is well-known that the number of internal degrees of freedom for a general $n$-dimensional unitary matrix is $(n-1)^2$, so there are nine internal degrees of freedom in this unitary matrix. After the possible phases are absorbed into external phases, the total number of degrees of freedom in $A$ and $B$ are eleven. Unitary conditions, i.e. orthonormality conditions eliminate four degrees of freedom. Thus, seven degrees of freedom remain. Two degrees of freedom have been removed from a non-degenerate case. It can be concluded that one degree of degeneracy corresponds to these two degrees of freedom. It is clear that a two-dimensional rotation matrix is a commutant in this case. It can be guessed that the other additional redundant parameter could come from a phase. This example can be easily generalized to $n$-dimensional case.\\\\
1. Degrees of freedom are contained only in $n-2$ vectors.\\
2. Each of $n-2$ vectors has $n$ real parameters, so total independent real parameters are $n(n-2)$.\\
3. All the phases from one of $n-2$ vectors can be absorbed into external phases, so $(n-1)(n-3)$ internal phases remain.\\
4. There are $2\times\frac{(n-2)(n-3)}{2}+n-2$ orthonormality conditions among $n-2$ vectors.\\
The total number of independent internal parameters is
\begin{equation}
n(n-2)+(n-1)(n-3)-(n-2)(n-3)-(n-2)=(n-1)^2-2.
\end{equation}
It can be seen that degeneracy is countable and identified with redundancy, i.e. one degree of degeneracy can be assigned to $\lambda_1=\lambda_2$ and corresponds to two degrees of redundancies (two redundant parameters) in a unitary matrix. Next, when one more eigenvalue is equal to $\lambda_1$ or $\lambda_2$, i.e. $\lambda_1=\lambda_2=\lambda_3$, one redundant column vector increases. As any arbitrary degrees of degeneracies for one eigenvalue are given, the number of independent parameters can be calculated. If $\Delta$ is defined the number of the same eigenvalues of one kind,
\begin{equation}\label{eq:internal degrees of freedom}
\begin{array}{ccl}
&&n(n-\Delta)+(n-1)(n-\Delta)-(n-1)-2\times\frac{(n-\Delta)(n-\Delta-1)}{2}-(n-\Delta)\\
&=&(n-1)^2-\Delta(\Delta-1).
\end{array}
\end{equation}
It can be noticed that the change $\Delta(\Delta-1)$ is the number of redundant parameters and is 2 times the number of possible pairs in the same eigenvalues. Thus, it can be concluded that two redundant parameters occur per one pair of eigenvalues. Now through the following argument this idea can be generalized. It is better to define degrees of degeneracy, the number of possible pairs in the same eigenvalues.
\begin{equation}
\text{Degrees of degeneracy}=\sum_i\frac{1}{2}\Delta_i(\Delta_i-1),
\end{equation}
where $\Delta_i$ is the number of the same eigenvalues of $i$ kind.
\begin{equation}
\text{The number of redundant parameters}=\sum_i\Delta_i(\Delta_i-1),
\end{equation}
If an additional distinct degeneracy is added, i.e. $\lambda_3=\lambda_4$.
\begin{equation}
U D^\prime=\Delta\lambda_{31}\left(\begin{array}{cccc}
\otimes&\otimes&a_1&b_1\\
\otimes&\otimes&a_2&b_2\\
\otimes&\otimes&a_3&b_3\\
\otimes&\otimes&a_4&b_4
\end{array}\right)
\left(\begin{array}{cccc}
0&0&0&0\\
0&0&0&0\\
0&0&1&0\\
0&0&0&1
\end{array}\right)
\end{equation}
and a density matrix is
\begin{equation}
\begin{array}{ccc}
\rho&=&\Delta\lambda_{31}\left(\begin{array}{cccc}
0&0&a_1&b_1\\
0&0&a_2&b_2\\
0&0&a_3&b_3\\
0&0&a_4&b_4
\end{array}\right)
\left(\begin{array}{cccc}
0&0&0&0\\
0&0&0&0\\
a^*_1&a^*_2&a^*_3&a^*_4\\
b^*_1&b^*_2&b^*_3&b^*_4
\end{array}\right)\\
&=&\Delta\lambda_{31}\left(\begin{array}{cccc}
|x_1|^2&x_1\cdot x^*_2&x_1\cdot x^*_3&x_1\cdot x^*_4\\
x_2\cdot x^*_1&|x_2|^2&x_2\cdot x^*_3&x_2\cdot x^*_4\\
x_3\cdot x^*_1&x_3\cdot x^*_2&|x_3|^2&x_3\cdot x^*_4\\
x_4\cdot x^*_1&x_4\cdot x^*_2&x_4\cdot x^*_3&|x_4|^2
\end{array}\right),
\end{array}
\end{equation}
where $x_i=(a_i,b_i)$. The last form has symmetries in a two-dimensional rotation and a common phase transformation,
\begin{equation}
\begin{array}{ccl}
(a_i,b_i)&\rightarrow & (e^{i\delta}a_i\cos\theta + b_i\sin\theta,-e^{i\delta}a_i\sin\theta +b_i\cos\theta)\\
&=&(a_i\cos\theta + e^{i\eta}b_i\sin\theta,-a_i\sin\theta + e^{i\eta}b_i\cos\theta).
\end{array}
\end{equation}
These symmetries eliminate two degrees of freedom, one rotation angle and one phase, in a unitary matrix. A more degenerate case, for instance $\lambda_3=\lambda_4=\lambda_5$, assigned to three degrees of degeneracy, three possible pairs, has six redundancies, i.e. three-dimensional rotation and three phases. This can be seen in Eq.(\ref{eq:UD2}) that there remain only one vector $B$ and one eigenvalue $\Delta\lambda_{41}$. All the phases on $B$  can be taken out to external phases $P_L$ and remaining degrees of freedom become three, which matches with Eq.(\ref{eq:internal degrees of freedom}), $(4-1)^2-3(3-1)=3$. 
\section{Commutant of an eigenvalue matrix}\label{Sec:Commutant of an eigenvalue matrix}
A commutant is a mathematical realization of symmetries between an eigenvalue matrix and a unitary matrix. Redundancy in a unitary matrix is just consequence of the symmetries and must be able to be expressed in terms of a commutant. That is, once a complete commutant for a density matrix is found, the redundant parmeters from a unitary matrix are expected to be factored out in a commutant form. As seen in the last section the degeneracies can be countable, identified with redundant parameters and thus should be expressed in terms of a commutant. As degrees of degeneracies increase one by one, a previous commutant should be changed, multiplied by another commutant. Thus, a unitary matrix itself should be formed as a product of the smallest possible commutants. It seems to be a rotation block associated with phases.\\
A commutant for a non-degenerate eigenvalue matrix is a diagonal phase matrix, so a phase matrix part in a unitary matrix, adjacent to an eigenvalue matrix is redundant regardless of degeneracy and must be removed. If eigenvalues are degenerate, the corresponding commutant has more degrees of freedom. For one degree of degeneracy, for example $\lambda_i=\lambda_j$, the corresponding commutant should include a product of a diagonal phase matrix and a two-dimensional rotation matrix. The most general possible commutant in this case is
\begin{equation}
D_n=\left(\begin{array}{cccc}
e^{i\delta_1}&0&\cdots&0\\
0&e^{i\delta_2}&\cdots&0\\
0&0&e^{i\delta_3}&0\\
0&0&\cdots&0\\
\vdots&\vdots&\vdots&\vdots\\
0&0&0&\cdots
\end{array}\right)
\left(\begin{array}{cccc}
1&0&\cdots&0\\
0&\ddots&\cdots&0\\
0&\ddots&\left(\begin{array}{cc}
c&s\\
-s&c\end{array}\right)&0\\
0&0&0&0\\
0&0&0&\cdots
\end{array}\right)
\left(\begin{array}{cccc}
e^{i\eta_1}&0&\cdots&0\\
0&e^{i\eta_2}&\cdots&0\\
0&0&e^{i\eta_3}&0\\
0&0&\cdots&0\\
\vdots&\vdots&\vdots&\vdots\\
0&0&0&\cdots
\end{array}\right).
\end{equation}
This is a possible unit for the smallest degeneracy. However, when a unitary matrix or a complete commutant is made by a product of such matrices, the number of independent phases have to be considered. In the above representation all the phases except one can be moved to either the left or the right to the rotation matrix,
\begin{equation}
D_n=\left(\begin{array}{cccc}
1&0&\cdots&0\\
0&\ddots&\cdots&0\\
0&\ddots&\left(\begin{array}{cc}
e^{i\delta^\prime_i}&0\\
0&1\end{array}\right)&0\\
0&0&0&0\\
0&0&0&\cdots
\end{array}\right)
\left(\begin{array}{cccc}
1&0&\cdots&0\\
0&\ddots&\cdots&0\\
0&\ddots&\left(\begin{array}{cc}
c&s\\
-s&c\end{array}\right)&0\\
0&0&0&0\\
0&0&0&\cdots
\end{array}\right)
\left(\begin{array}{cccc}
e^{i\eta^\prime_1}&0&\cdots&0\\
0&e^{i\eta^\prime_2}&\cdots&0\\
0&0&e^{i\eta^\prime_3}&0\\
0&0&\cdots&0\\
\vdots&\vdots&\vdots&\vdots\\
0&0&0&\cdots
\end{array}\right).
\end{equation}
The phase matrix on the right to the rotation matrix can be again moved further leaving only one phase when another unit is multiplied to the right. A unitary matrix can be made by multiplying all the possible block each of which consists of one left phase, a two-dimensional rotation and a general phase matrix on the right. A commutant can be also made in the same way. Define a convenient block, $W$
\begin{equation}
W=\left(\begin{array}{cc}
e^{i\delta}&0\\
0&1\end{array}\right)
\left(\begin{array}{cc}
c&s\\
-s&c\end{array}\right).
\end{equation}
A unitary matrix can be written as a product of these $n(n-1)/2$ units $W_i$ and a general phase matrix $Q_n$.
\begin{equation}
U_n=W_1W_2\cdots W_{n(n-1)/2}Q_n,
\end{equation}
where
\begin{equation}
W_i=\left(\begin{array}{cccc}
1&0&\cdots&0\\
0&\ddots&\cdots&0\\
0&\ddots&W
&0\\
\vdots&\vdots&\vdots&\vdots\\
0&0&0&\cdots
\end{array}\right),~~~
Q_n=\left(\begin{array}{cccc}
e^{i\eta_1}&0&\cdots&0\\
0&e^{i\eta_2}&\cdots&0\\
0&0&e^{i\eta_3}&0\\
0&0&\cdots&0\\
\vdots&\vdots&\vdots&\vdots\\
0&0&0&\cdots
\end{array}\right).
\end{equation}
Let this representation be called ``the one phase-one rotation'' representation. It is known that the total number of independent parameters for a unitary matrix for $n$-dimension is $n^2$. It can be verified that the number of parameters in this representation also matches with it.
\begin{equation}
\frac{n(n-1)}{2}\times 2+n=n^2.
\end{equation}

\section{Equivalent angular representations}\label{Sec:Equivalent angular representations}
\subsection{Surjectivity}
It is practically important to know which angular representations among given representations of unitary matrices with rotations and phases are equivalent to a general description. Before considering this problem it is worthwhile to clarify whether an angular representation is onto, i.e. surjective to the canonical coordinates of the first kind, i.e. exponential map $U=e^X$, where $X$ is a general $n$-dimensional anti-hermitian matrix. By using the Baker\--Campbell\--Hausdorff formula it is obvious that the domain of the representation $U=e^X$ is greater than or equal to that of the representation $e^{X_1}\cdots e^{X_r}$, where $X=X_1+\cdots+X_r=X$ is composed of $X_i(\theta_\alpha,\cdots)$ parametrized with $(\theta_\alpha, \cdots)$.\\
In the same way it can be proved that all the group elements in the form $U=e^X$ can be written in a from $e^{X_1}\cdots e^{X_r}$, that is, both of the representations are equivalent. Assume that there is an element $e^X$, where $X=tX_1+\cdots+tX_r$ which could not be written in some decomposition form $e^{tX^\prime_1}\cdots e^{tX^\prime_r}$, where $X^\prime_i=X_i(\theta^\prime_\alpha,\cdots)$. By using the Baker\--Campbell\--Hausdorff formula,
\begin{equation}
e^{tX_1+t^2\delta X_1}\cdots e^{tX_r+t^2\delta X_r}=e^{tX_1+\cdots +tX_r +t^2(\delta X_1+\delta X_2+\cdots+\delta X_r)+O(t^2)+O(t^3)},
\end{equation}
where $\delta X_i$ has not been specified yet. To eliminate $t^2$ order terms in the exponential in the right hand side, choose $t^2(\delta X_1+\cdots+\delta X_r)$ as $-O(t^2)$, which is coming from the commutators among $tX_i$, independent of $\delta X_i$. This implies that $e^{tX_1+\cdots+tX_r}$ is equal to $e^{tX_1+t^2\delta X_1}\cdots e^{tX_r+t^2\delta X_r}$ up to $O(t^3)$. Again, $t^3\delta X^\prime_i$ can be included in the exponential in the left hand side to eliminate $t^3$ order. In this way, all the higher order differences can be removed up to any order. It means that $e^{tX_1+\cdots+tX_r}$ can be expressed in a from $e^{tX^\prime_1}\cdots e^{tX^\prime_r}$.
\begin{equation}
e^{tX_1+t^2\delta X_1+t^3\delta X^\prime_1}\cdots e^{tX_r+t^2\delta X_r+t^3\delta X^\prime_r}=e^{tX_1+\cdots +tX_r +t^3(\delta X^\prime_1+\delta X^\prime_2+\cdots+\delta X^\prime_r)+O(t^3)}.
\end{equation}
Thus, the previous assumption is not correct. Therefore, if $X=X_1+\cdots+X_r$ is a general anti-hermitian matrix, the exponential map $e^{X_1}\cdots e^{X_r}$ is onto the representation in the canonical coordinates of the first kind $e^X$.

\subsection{Phase transformed rotation matrix}\label{Sec:Phase transformed rotation matrix}
 As it has been just proven that any decomposition from a general anti-hermitian matrix $X$ is equivalent to a representation in the canonical coordinates of the first kind. It is well-known that a three-dimensional unitary matrix can be represented with three rotations, one internal and five external phases. It is called the KM parametrization in a particle physics \cite{Kobayashi:1973fv}. It can be also generalized to $n$-dimension. This representation is suitable when external phases are not physical and absorbed into fermion fields. Here starting with a possible simple decomposition from the representation in the canonical coordinates of the first kind, it will be shown that it can reduced to the representation, the one phase-one rotation-representation, defined in the section \ref{Sec:Commutant of an eigenvalue matrix}. It is a convenient representation for a degenerate density matrix. It will be shown how phases can be moved to another places using diagrams in the last subsection \ref{Sec:Phase manipulations}. A general anti-hermitian matrix $X$ in $n$-dimension is

\begin{equation}
X=\left(\begin{array}{ccccc}
i\alpha_1&z_{12}&\cdots&z_{1,n-1}&z_{1,n}\\
-z^*_{12}&i\alpha_2&z_{23}&\cdots&z_{2,n}\\
\vdots&\vdots&\ddots&\vdots&\vdots\\
-z^*_{1,n-1}&-z^*_{1,n-1}&\cdots&i\alpha_{n-1}&z_{n-1,n}\\
-z^*_{1,n}&-z^*_{2,n}&\cdots&-z^*_{n-1,n}&i\alpha_n
\end{array}\right).
\end{equation}
Let $X$ be decomposed as follows.
\begin{equation}
X_{1,1}=\left(\begin{array}{ccccc}
i\alpha_1&0&\cdots&0&0\\
0&0&0&0&0\\
\vdots&\vdots&\ddots&\vdots&\vdots\\
0&0&0&0&0\\
0&0&0&0&0
\end{array}\right),~~
X_{2,2}=\left(\begin{array}{ccccc}
0&0&\cdots&0&0\\
0&i\alpha_2&0&0&0\\
\vdots&\vdots&\ddots&\vdots&\vdots\\
0&0&0&0&0\\
0&0&0&0&0
\end{array}\right),\cdots
X_{n,n}=\left(\begin{array}{ccccc}
0&0&\cdots&0&0\\
0&0&0&0&0\\
\vdots&\vdots&\ddots&\vdots&\vdots\\
0&0&0&0&0\\
0&0&0&0&i\alpha_{n}
\end{array}\right),
\end{equation}

\begin{equation}
X_{12}=\left(\begin{array}{ccccc}
0&z_{12}&\cdots&0&0\\
-z^*_{12}&0&0&0&0\\
\vdots&\vdots&\ddots&\vdots&\vdots\\
0&0&0&0&0\\
0&0&0&0&0
\end{array}\right),~~
X_{13}=\left(\begin{array}{ccccc}
0&0&z_{13}&\cdots&0\\
0&0&0&0&0\\
-z^*_{13}&0&\ddots&\vdots&\vdots\\
0&0&0&0&0\\
0&0&0&0&0
\end{array}\right),\cdots
X_{n-1, n}=\left(\begin{array}{ccccc}
0&0&\cdots&0&0\\
0&0&0&0&0\\
\vdots&\vdots&\ddots&\vdots&\vdots\\
0&0&0&0&z_{n-1,n}\\
0&0&0&-z^*_{n-1,n}&0
\end{array}\right).\nonumber
\end{equation}
A vector $z_{ij}$ can be written $e^{i\theta_{ij}}|z_{ij}|$, so the matrix $e^{X_{ij}}$ can be expressed,
\begin{equation}
e^{X_{ij}}=e^{X_{i,i}(\theta_{ij})}e^{X_{ij}(|z_{ij}|)}e^{X_{i,i}^\dagger(\theta_{ij})}
=e^{X_{j,j}(-\theta_{ij})}e^{X_{ij}(|z_{ij}|)}e^{X_{j,j}^\dagger(-\theta_{ij})},
\end{equation}
In other words, $e^{X_{ij}}=P_{i}(\theta_{ij})R_{ij}(|z_{ij}|)P^\dagger_i(\theta_{ij})=P_{j}(-\theta_{ij})R_{ij}(|z_{ij}|)P^\dagger_j(-\theta_{ij})$. $X_{ij}$ matrix has one rotation and its one associated phase degrees of freedom. A unitary matrix in this decomposition is given, 
\begin{equation}
U_n=e^{X_{12}}e^{X_{23}}\cdots e^{X_{n-1, n}}Q_n
=P_1R_{12}P^\dagger_1 P_2R_{23}P^\dagger_2\cdots P_{n-1}R_{n-1,n}P^\dagger_{n-1}Q_n.
\end{equation}
where $Q_n$ is a general phase matrix in $n$-dimension. Note that $Q_n$ can be placed in anywhere, but for the purpose of dealing with a degeneracy it is more convenient to put on the right side. Let us call this representation,  ``the  phase adjoint rotation representation'' here.
In an angular representation the number of internal and external phases is
\begin{equation}
\frac{(n-1)(n-2)}{2}+2n-1=\frac{1}{2}n(n+1).
\end{equation}
The number of independent phases in a usual angular representation is exactly the same as that of phases in this representation. The number of phases from $X_{i,i}$ generators is $n$ and the number of phases from each rotation is $\frac{1}{2}n(n-1)$.
\begin{equation}
\frac{1}{2}n(n-1)+n=\frac{1}{2}n(n+1).
\end{equation}
The following subsection will show how this representation is transformed to the one phase-one rotation representation with an example for $n=3$  
\subsection{Degenerate density matrix for $n=3$}\label{Sec:Degenerate density matrix for n=3}
This section presents an example of a degenerate density matrix for $n=3$ parametrized with a certain choice of a phase representation of a unitary matrix, the phase adjoint rotation representation, defined in the last section \ref{Sec:Phase transformed rotation matrix}, and spherical polar coordinates for a eigenvalue matrix in the spectral representation, $\rho_3=U_3D_3U^\dagger_3$.
The unitary matrix representation here consists of $n(n-1)/2=3$ adjoint rotation matrices $W_{ij}=P_iR_{ij}P^\dagger_i$, with each of $P_i$ having one independent phase and a general phase matrix $Q_n$. All the representations in different orders of matrices are equivalent. 
To satisfy the non-negativity of eigenvalues and the normalized trace condition spherical polar coordinates are chosen. For convenience a general phase matrix $Q_n$ are placed on the right in a unitary matrix.
\begin{equation}\label{U3}
U_3=W_{31}W_{23}W_{12}Q_3.
\end{equation}
The associated phase matrices with rotation matrices are
\begin{equation}
P_1=\left(\begin{array}{ccc}
e^{i\delta_{1}}&0&0\\
0&1&0\\
0&0&1
\end{array}\right),~~
P_2=\left(\begin{array}{ccc}
1&0&0\\
0&e^{i\delta_{2}}&0\\
0&0&1
\end{array}\right),~~
P_3=\left(\begin{array}{ccc}
1&0&0\\
0&1&0\\
0&0&e^{i\delta_{3}}
\end{array}\right),
\end{equation}
where the parameter set for $\delta_i$ is 
\begin{equation}
S_p=\{\delta_i\in R: 0\leq\delta_i< 2\pi, (i=1,2,3) \}.
\end{equation}
In addition to those there are $n$ independent phases necessary for completion. 
\begin{equation}
Q_3=\left(\begin{array}{ccc}
e^{i\eta_{1}}&0&0\\
0&e^{i\eta_{2}}&0\\
0&0&e^{i\eta_{3}}
\end{array}\right),
\end{equation}
where the parameter set for $\eta_i$ is 
\begin{equation}
S_Q=\{\eta_i\in R: 0\leq\eta_i< 2\pi, (i=1,2,3) \}.
\end{equation}
$n(n-1)/2$ possible rotations are
\begin{equation}
R_{12}=\left(\begin{array}{ccc}
c_{12}&s_{12}&0\\
-s_{12}&c_{12}&0\\
0&0&1
\end{array}\right),~~
R_{23}=\left(\begin{array}{ccc}
1&0&0\\
0&c_{23}&s_{23}\\
0&-s_{23}&c_{23}
\end{array}\right),~~ \text{and}~
R_{31}=\left(\begin{array}{ccc}
c_{31}&0&s_{31}\\
0&1&0\\
-s_{31}&0&c_{31}
\end{array}\right),
\end{equation}
where the parameter set for $(\theta_{12}, \theta_{23}, \theta_{31})$ is
\begin{equation}
S_R=\{(\theta_{12}, \theta_{23}, \theta_{31}\in R^3:0\leq \theta_{12}, \theta_{23},\theta_{31}\leq \pi/2 \}.
\end{equation}
One can check that the ranges of rotation angles $0\leq \theta_{ij}\leq \pi/2$ guarantees the mapping to be one-to-one. The phase $e^{i\delta_i}$ at $\delta_i=\pi$ amounts to change of signs in some elements in rotation matrices. $W_{ij}$ has the left phase to rotation matrix $R_{ij}$. This phase together with  phases from $W_{ik}$ and $W_{jk}$ which can be brought to the right side to the $R_{ij}$, can change a sign of either cosine or sine in $R_{ij}$. With one associated phase to a rotation matrix, a sign of only one column or row can change, but together with another phase from the above relevant $W$ matrices on the other side a sign of either cosine or sine can be changed. Therefore, the ranges of rotation angles, $0\leq\theta_{ij}\leq \pi/2$ is necessary and sufficient to cover all possible values for rotations. 
A unitary matrix can be explicitly expressed from Eq.(\ref{U3}),
\begin{equation}
\begin{array}{ccl}
U_3=&&\left(\begin{array}{ccc}
1&0&0\\
0&1&0\\
0&0&e^{i\delta_{3}}
\end{array}\right)
\left(\begin{array}{ccc}
c_{31}&0&s_{31}\\
0&1&0\\
-s_{31}&0&c_{31}
\end{array}\right)
\left(\begin{array}{ccc}
1&0&0\\
0&e^{i\delta_{2}}&0\\
0&0&e^{-i\delta_{3}}
\end{array}\right)
\left(\begin{array}{ccc}
1&0&0\\
0&c_{23}&s_{23}\\
0&-s_{23}&c_{23}
\end{array}\right)\\
&\times &\left(\begin{array}{ccc}
e^{i\delta_{1}}&0&0\\
0&e^{-i\delta_{2}}&0\\
0&0&1
\end{array}\right)
\left(\begin{array}{ccc}
c_{12}&s_{12}&0\\
-s_{12}&c_{12}&0\\
0&0&1
\end{array}\right)
\left(\begin{array}{ccc}
e^{-i\delta_{1}}&0&0\\
0&1&0\\
0&0&1
\end{array}\right)
\left(\begin{array}{ccc}
e^{i\eta_{1}}&0&0\\
0&e^{i\eta_{2}}&0\\
0&0&e^{i\eta_{3}}
\end{array}\right).
\end{array}
\end{equation}
By the following observation, a phase can pass through a rotation matrix. 
\begin{equation}
\left(\begin{array}{cc}
e^{i\delta_{1}}&0\\
0&e^{i\delta_{2}}
\end{array}\right)
\left(\begin{array}{cc}
c&s\\
-s&c
\end{array}\right)
=\left(\begin{array}{cc}
e^{i(\delta_{1}-\delta_{2})}&0\\
0&1
\end{array}\right)
\left(\begin{array}{cc}
c&s\\
-s&c
\end{array}\right)
\left(\begin{array}{cc}
e^{i\delta_{2}}&0\\
0&e^{i\delta_{2}}
\end{array}\right),
\end{equation}
A unitary matrix can be rearranged as a product of the one rotation-one right phase blocks and $n$ phases.
\begin{equation}
\begin{array}{ccl}
U_3=&&\left(\begin{array}{ccc}
1&0&0\\
0&1&0\\
0&0&e^{i\delta_{3}}
\end{array}\right)
\left(\begin{array}{ccc}
c_{31}&0&s_{31}\\
0&1&0\\
-s_{31}&0&c_{31}
\end{array}\right)
\left(\begin{array}{ccc}
1&0&0\\
0&e^{i(\delta_{2}+\delta_{3})}&0\\
0&0&1
\end{array}\right)
\left(\begin{array}{ccc}
1&0&0\\
0&c_{23}&s_{23}\\
0&-s_{23}&c_{23}
\end{array}\right)
\left(\begin{array}{ccc}
1&0&0\\
0&e^{-i(\delta_{2}+\delta_{3})}&0\\
0&0&1
\end{array}\right)\\
&\times &
\left(\begin{array}{ccc}
1&0&0\\
0&e^{i\delta_{2}}&0\\
0&0&e^{i\delta_{3}}
\end{array}\right)
\left(\begin{array}{ccc}
e^{i\delta_{1}}&0&0\\
0&e^{-i\delta_{2}}&0\\
0&0&1
\end{array}\right)
\left(\begin{array}{ccc}
c_{12}&s_{12}&0\\
-s_{12}&c_{12}&0\\
0&0&1
\end{array}\right)
\left(\begin{array}{ccc}
e^{-i\delta_{1}}&0&0\\
0&1&0\\
0&0&1
\end{array}\right)
\left(\begin{array}{ccc}
e^{i\eta_{1}}&0&0\\
0&e^{i\eta_{2}}&0\\
0&0&e^{i\eta_{3}}
\end{array}\right)\\
=&&\left(\begin{array}{ccc}
1&0&0\\
0&1&0\\
0&0&e^{i\delta_{3}}
\end{array}\right)
\left(\begin{array}{ccc}
c_{31}&0&s_{31}\\
0&1&0\\
-s_{31}&0&c_{31}
\end{array}\right)
\left(\begin{array}{ccc}
1&0&0\\
0&e^{i(\delta_{2}+\delta_{3})}&0\\
0&0&1
\end{array}\right)
\left(\begin{array}{ccc}
1&0&0\\
0&c_{23}&s_{23}\\
0&-s_{23}&c_{23}
\end{array}\right)\\
&\times &
\left(\begin{array}{ccc}
e^{i(\delta_{1}+\delta_{2}+\delta_{3})}&0&0\\
0&1&0\\
0&0&1
\end{array}\right)
\left(\begin{array}{ccc}
c_{12}&s_{12}&0\\
-s_{12}&c_{12}&0\\
0&0&1
\end{array}\right)
\left(\begin{array}{ccc}
e^{-i(\delta_{1}+\delta_{2}+\delta_{3})}&0&0\\
0&e^{i(\delta_{2}+\delta_{3})}&0\\
0&0&e^{i\delta_{3}}
\end{array}\right)
\left(\begin{array}{ccc}
e^{i\eta_{1}}&0&0\\
0&e^{i\eta_{2}}&0\\
0&0&e^{i\eta_{3}}
\end{array}\right).
\end{array}
\end{equation}
It can be realized that a unitary matrix is just a product of such blocks, $Y_i$ and a phase matrix, $Q_3$.
\begin{equation}
U_3=Y_3Y_2Y_1Q_3,
\end{equation}
where
\begin{equation}
Y_3=P_3R_{31}=\left(\begin{array}{ccc}
1&0&0\\
0&1&0\\
0&0&e^{i\delta_{3}}
\end{array}\right)
\left(\begin{array}{ccc}
c_{31}&0&s_{31}\\
0&1&0\\
-s_{31}&0&c_{31}
\end{array}\right),
\end{equation}
\begin{equation}
Y_2=P_2R_{23}
=\left(\begin{array}{ccc}
1&0&0\\
0&e^{i\delta_{2}}&0\\
0&0&1
\end{array}\right)
\left(\begin{array}{ccc}
1&0&0\\
0&c_{23}&s_{23}\\
0&-s_{23}&c_{23}
\end{array}\right)
\end{equation}
and
\begin{equation}
Y_1=P_1R_{12}
=\left(\begin{array}{ccc}
e^{i\delta_{1}}&0&0\\
0&1&0\\
0&0&1
\end{array}\right)
\left(\begin{array}{ccc}
c_{12}&s_{12}&0\\
-s_{12}&c_{12}&0\\
0&0&1
\end{array}\right).
\end{equation}
A density matrix in the spectral representation is 
\begin{equation}
\rho_3=U_3D_3U^\dagger_3.
\end{equation}
Certain parametrizations for a diagonal eigenvalue matrix automatically satisfy the trace and the non-negativity conditions. Such parametrizations are not so relevant for $n=3$ case, but as general cases considered those parametrizations may be useful. Let use spherical polar coordinates \cite{Boya:1998nq, Tilma:2002kf}.

\begin{equation}
D_3=\left(\begin{array}{ccc}
\sin^2\theta\sin^2\phi&0&0\\
0&\sin^2\theta\cos^2\phi&0\\
0&0&\cos^2\theta
\end{array}\right).
\end{equation}
where the parameter set for $(\theta, \phi)$ is
\begin{equation}
S_D=\{(\theta, \phi)\in R^2:0\leq \theta\leq \pi/2,0\leq \phi\leq \pi/2 \}.
\end{equation}
The total parameter space for a density matrix for $n=3$ is
\begin{equation}
S=\{(\delta_i,\theta_{12},\theta_{23}, \theta_{31}, \theta, \phi)\in R^8:0\leq\delta_i\leq 2\pi, 0\leq \theta_{12},\theta_{23}, \theta_{31}, \theta, \phi \leq \pi/2, (i=1,2,3) \}.
\end{equation}
This parametrization for a eigenvalue matrix can be extended to $n$-dimensional case using $n$-sphere polar coordinates. If only $k$ eigenvalues are distinct due to degeneracy, instead of using $n$-sphere polar coordinates $k$-sphere polar coordinates are used. The same averaged squared component is given to the the same eigenvalues. For example, in the examples below two dimensional polar coordinates are used and $\lambda_1=\lambda_2$ are the averaged squared component $\frac{1}{2}\sin^2\theta$. Similarly, for $\lambda_2=\lambda_3$,  $\lambda_2=\lambda_3=\frac{1}{2}\cos^2\theta$.\\
i) Case $\lambda_1=\lambda_2$.
\begin{equation}
\rho_{12}=Y_3Y_2Y_1QD_3Q^\dagger Y^\dagger_1Y^\dagger_2Y^\dagger_3
=Y_3Y_2D_3Y^\dagger_2Y^\dagger_3,
\end{equation}
\begin{equation}
\begin{array}{ccl}
\rho_{12}=&&\left(\begin{array}{ccc}
1&0&0\\
0&1&0\\
0&0&e^{i\delta_{3}}
\end{array}\right)
\left(\begin{array}{ccc}
c_{31}&0&s_{31}\\
0&1&0\\
-s_{31}&0&c_{31}
\end{array}\right)
\left(\begin{array}{ccc}
1&0&0\\
0&e^{i\delta_{2}}&0\\
0&0&1
\end{array}\right)
\left(\begin{array}{ccc}
1&0&0\\
0&c_{23}&s_{23}\\
0&-s_{23}&c_{23}
\end{array}\right)
\left(\begin{array}{ccc}
s^2_\theta /2&0&0\\
0&s^2_\theta /2&0\\
0&0&c^2_\theta
\end{array}\right)\\
&&\times
\left(\begin{array}{ccc}
1&0&0\\
0&c_{23}&-s_{23}\\
0&s_{23}&c_{23}
\end{array}\right)
\left(\begin{array}{ccc}
1&0&0\\
0&e^{-i\delta_{2}}&0\\
0&0&1
\end{array}\right)
\left(\begin{array}{ccc}
c_{31}&0&-s_{31}\\
0&1&0\\
s_{31}&0&c_{31}
\end{array}\right)
\left(\begin{array}{ccc}
1&0&0\\
0&1&0\\
0&0&e^{-i\delta_{3}}
\end{array}\right).
\end{array}
\end{equation}
ii) Case $\lambda_2=\lambda_3$.
\begin{equation}
\rho_{12}=Y_3Y_1Y_2QD_3Q^\dagger Y^\dagger_2Y^\dagger_1Y^\dagger_3
=Y_3Y_1D_3Y^\dagger_1Y^\dagger_3,
\end{equation}
\begin{equation}
\begin{array}{ccl}
\rho_{23}=&&\left(\begin{array}{ccc}
1&0&0\\
0&1&0\\
0&0&e^{i\delta_{3}}
\end{array}\right)
\left(\begin{array}{ccc}
c_{31}&0&s_{31}\\
0&1&0\\
-s_{31}&0&c_{31}
\end{array}\right)
\left(\begin{array}{ccc}
e^{-i\delta_{1}}&0&0\\
0&1&0\\
0&0&1
\end{array}\right)
\left(\begin{array}{ccc}
c_{12}&s_{12}&0\\
-s_{12}&c_{12}&0\\
0&0&1
\end{array}\right)
\left(\begin{array}{ccc}
s^2_\theta&0&0\\
0&c^2_\theta/2 &0\\
0&0& c^2_\theta/2
\end{array}\right)\\
&&\times
\left(\begin{array}{ccc}
c_{12}&-s_{12}&0\\
s_{12}&c_{12}&0\\
0&0&1
\end{array}\right)
\left(\begin{array}{ccc}
e^{-i\delta_{1}}&0&0\\
0&1&0\\
0&0&1
\end{array}\right)
\left(\begin{array}{ccc}
c_{31}&0&-s_{31}\\
0&1&0\\
s_{31}&0&c_{31}
\end{array}\right)
\left(\begin{array}{ccc}
1&0&0\\
0&1&0\\
0&0&e^{-i\delta_{3}}
\end{array}\right).
\end{array}
\end{equation}
This example can be easily extended to $n$-dimensional unitary matrix with $n$-dimensional spherical coordinates for a parametrization of an eigenvalue matrix. 
\subsection{Phase manipulations}\label{Sec:Phase manipulations}
For practical purposes it would be often useful if there is an easy way to check whether a certain phase representation is a general description of a unitary matrix. As seen the one phase-one rotation representation is a general description of a unitary matrix. Here a simple diagram is introduced to show that a product of all possible rotation matrix with general phase matrices inserted between them is also equivalent, changing a phase configuration.
\begin{equation}\label{eq:general phase representation}
U=P_LR_{12}P_1R_{23}P_2\cdots R_{n,n-1}P_R=Y_1Y_2\cdots Y_{n(n-1)/2}P^\prime_R,
\end{equation}
where $Y_i=\tilde{P}_iR_{ij}$, $\tilde{P}_i$ has a phase matrix with one phase in $(i,i)$ place, $P_i$, $P_L$ and $P_R$  are general diagonal phase matrices and $R_{ij}$ are rotation matrices. The following simplified diagrams, in Tables [\ref{tb:1}, \ref{tb:2}], are convenient to transform one phase representation to another. 
\begin{table}[!ht]
\centering
    $\left(\begin{array}{cc}
\cos\theta_{ij}&\sin\theta_{ij}\\
-\sin\theta_{ij}&\cos\theta_{ij}\\
\end{array}\right)=~$\begin{tabular}{|l|}
       \hline
          $i$\\ \hline
         $j$\\ \hline
    \end{tabular}~,~~
     $\left(\begin{array}{cc}
e^{i\delta_i}&0\\
0&e^{i\delta_j}\\
\end{array}\right)=~$\begin{tabular}{|l|}
       \hline
          $\textcircled{i}$\\ \hline
         $\textcircled{j}$\\ \hline
    \end{tabular}
    \caption{Rotation blocks and phase matrices}\label{tb:1}
    \end{table}
\begin{table}[!ht]
\centering
    $\left(\begin{array}{cc}
e^{i\delta_i}&0\\
0&e^{i\delta_j}\\
\end{array}\right)\left(\begin{array}{cc}
\cos\theta_{ij}&\sin\theta_{ij}\\
-\sin\theta_{ij}&\cos\theta_{ij}\\
\end{array}\right)
\left(\begin{array}{cc}
e^{i\delta_k}&0\\
0&e^{i\delta_l}\\
\end{array}\right)=~$\begin{tabular}{|l|l|l|}
       \hline
          $\textcircled{i}$ &$i$&$\textcircled{k}$\\ \hline
         $\textcircled{j}$ &$j$&$\textcircled{l}$\\ \hline  
    \end{tabular}
    \caption{a phase-rotation-phase product}\label{tb:2}
    \end{table}
The first example shows that the above representation can be reduced to the KM parametrization\cite{Kobayashi:1973fv}. The key point in this manipulation is to realize that a general $U(2)$ matrix must be parametrized with not four but three independent phases and one rotation. From four parameters in Eq.(\ref{3phaseU(2)}) can be reduced to three independent phases because $U(1)$ phase can commute with any matrix. Also, places for three independent phases can be free to be changed out of four places.
\begin{equation}\label{3phaseU(2)}
\left(\begin{array}{cc}
e^{i\delta_i} &0\\
0&e^{i\delta_j}\\
\end{array}\right)
 \left(\begin{array}{cc}
c&s\\
-s&c\\
\end{array}\right)
\left(\begin{array}{cc}
e^{i\eta_i} &0\\
0&e^{i\eta_j}
\end{array}\right)
\Rightarrow
\left(\begin{array}{cc}
e^{i(\delta_i-\delta_j)} &0\\
0&1\\
\end{array}\right)
 \left(\begin{array}{cc}
c&s\\
-s&c\\
\end{array}\right)
\left(\begin{array}{cc}
e^{i(\eta_i+\delta_j)} &0\\
0&e^{i(\eta_j+\delta_j)}
\end{array}\right)
\end{equation}
Tables [\ref{tb:3}, \ref{tb:4}, \ref{tb:5}] show how to manipulate phases in the general phase representation Eq.(\ref{eq:general phase representation}) to the KM parametrization \cite{Kobayashi:1973fv}. 
\begin{table}[!ht]
    \begin{tabular}{|c|c|c|c|c|c|c|}
       \hline
         $\textcircled{L}_1$&  & $\textcircled{1}_L$&1&$\textcircled{1}_C$ &1 &$\textcircled{R}_1$\\ \hline
         $\textcircled{L}_2$ &2 &$\textcircled{2}_C$&2 &$\textcircled{2}_R$&&$\textcircled{R}_2$\\ \hline
         $\textcircled{L}_3$ & 3 &$\textcircled{3}_R$&&$\textcircled{3}_L$&3&$\textcircled{R}_3$\\
        \hline
    \end{tabular}~~$\rightarrow$
     \begin{tabular}{|c|c|c|c|c|c|c|}
        \hline
        &  & $\textcircled{L}_1\textcircled{1}_L$&1&$\textcircled{1}_C$&1 &$\textcircled{R}_1$\\ \hline
         $\textcircled{L}_2$ &2 &$\textcircled{2}_C$&2 &$\textcircled{2}_R\textcircled{R}_2$&&\\ \hline
         $\textcircled{L}_3$ & 3 &&&$\textcircled{3}_R\textcircled{3}_L$&3&$\textcircled{R}_3$\\
        \hline
    \end{tabular}
    \caption{Combining commuting phases}\label{tb:3}
    \end{table}
    \begin{table}[!ht]
    \begin{tabular}{|c|c|c|c|c|c|c|}
        \hline
        &  & $\textcircled{L}_1\textcircled{1}_L$&1&$\textcircled{1}_C$&1 &$\textcircled{R}_1$\\ \hline
         $\textcircled{L}_2$ &2 &$\textcircled{2}_C$&2 &$\textcircled{2}_R\textcircled{R}_2$&&\\ \hline
         $\textcircled{L}_3$ & 3 &&&$\textcircled{3}_R\textcircled{3}_L$&3&$\textcircled{R}_3$\\
        \hline
    \end{tabular}~~$\rightarrow$
     \begin{tabular}{|c|c|c|c|c|c|c|}
        \hline
        &  & $\textcircled{L}_1\textcircled{1}_L$&1&$\textcircled{1}_C/\textcircled{3}_R\textcircled{3}_L$&1 &$\textcircled{R}_1\textcircled{3}_R\textcircled{3}_L$\\ \hline
         $\textcircled{L}_2$ &2 &$\textcircled{2}_C$&2 &$\textcircled{2}_R\textcircled{R}_2/\textcircled{3}_R\textcircled{3}_L$&&$\textcircled{3}_R\textcircled{3}_L$\\ \hline
         $\textcircled{L}_3$ & 3 &&&&3&$\textcircled{R}_3\textcircled{3}_R\textcircled{3}_L$\\
        \hline
    \end{tabular}
    \caption{One phase can be removed from four phases around (3,1) rotation.}\label{tb:4}
\end{table}
\begin{table}[!ht]
    \begin{tabular}{|c|c|c|c|c|c|c|}
    \hline
        &  & $\textcircled{1}^\prime_L$&1&$\textcircled{1}^\prime_C$&1 &$\textcircled{R}_1^\prime$\\ \hline
         $\textcircled{L}_2$ &2 &$\textcircled{2}_C$&2 &$\textcircled{2}_R^\prime$&&\\ \hline
         $\textcircled{L}_3$ & 3 &&&&3&$\textcircled{R}_3^\prime$\\
        \hline
    \end{tabular}
    $\rightarrow$
     \begin{tabular}{|c|c|c|c|c|c|c|}
       \hline
        &  & $\textcircled{1}^\prime_L/\textcircled{2}_R$&1&$\textcircled{1}^\prime_C/\textcircled{2}_R$&1 &$\textcircled{R}_1^\prime$\\ \hline
         $\textcircled{L}_2$ &2 &&2 &$\textcircled{2}_R^\prime/\textcircled{2}_C$&&\\ \hline
         $\textcircled{L}_3$ & 3 &&&&3&$\textcircled{R}_3^\prime$\\
        \hline
    \end{tabular}
     \begin{center}
    $\rightarrow$
     \begin{tabular}{|c|c|c|c|c|c|c|}
       \hline
       $\textcircled{1}^\prime_L/\textcircled{2}_R$ &  & &1&$\textcircled{1}^\prime_C/\textcircled{2}_R$&1 &$\textcircled{R}_1^\prime$\\ \hline
         $\textcircled{L}_2$ &2 &&2 &&&$\textcircled{2}_R^\prime/\textcircled{2}_C$\\ \hline
         $\textcircled{L}_3$ & 3 &&&&3&$\textcircled{R}_3^\prime$\\
        \hline
    \end{tabular}
    \caption{One phase can be removed from four phases around (1,2) rotation.}\label{tb:5}
 \end{center}
\end{table}
In a similar way the phase adjoint rotation representation defined in the subsection \ref{Sec:Phase transformed rotation matrix} can be shown to other representations, the KM, the one phase-one rotation representation defined in the section \ref{Sec:Commutant of an eigenvalue matrix} and so on.
\section{Conclusion}
It has been presented that degeneracies are quantified as the number of pairs of the same eigenvalues and identified with redundant parameters in a unitary matrix. Redundant parameters due to degenerate eigenvalues in an $n$-dimensional density matrix can be conveniently eliminated after a commutant is factored out from a density matrix. It has been found that the one phase-one rotation representation in the section \ref{Sec:Commutant of an eigenvalue matrix} is a convenient representation to factor a unitary matrix with commutant units. It is a suitable way to factor out redundancies in a unitary matrix. In practice, there are many different possible angular representations just by changing phase configurations. It is often useful to figure out whether a given representation is general. A simple manipulation has been shown for transforming one representation to another.
\begin{acknowledgments}
I would like to express my appreciation to Prof. Erwin Bruning, who guided me with a lot of essential comments and advice in this research.  
\end{acknowledgments}

\end{document}